\documentclass[useAMS,usenatbib,usegraphicx]{mn2e}
\usepackage{natbib}
\usepackage{pdfpages}
\usepackage{graphicx}
\usepackage{titlesec}
\titlespacing*{\section}{0pt}{1.1\baselineskip}{\baselineskip}
\newcommand{\gae}{\lower 2pt \hbox{$\, \buildrel {\scriptstyle >}\over {\scriptstyle
\sim}\,$}}
\newcommand{\lae}{\lower 2pt \hbox{$\, \buildrel {\scriptstyle <}\over {\scriptstyle
\sim}\,$}}

\title[ULTRASPEC on the 2.4-m TNT]{ULTRASPEC: a high-speed imaging
  photometer on the 2.4-m Thai National Telescope}

\author[V. S. Dhillon et al.]
{V. S. Dhillon,$^1$\thanks{E-mail: vik.dhillon@sheffield.ac.uk}
T. R. Marsh,$^2$\thanks{E-mail: t.r.marsh@warwick.ac.uk}
D. C. Atkinson,$^3$
N. Bezawada,$^3$
M. C. P. Bours,$^2$ 
\newauthor
C. M. Copperwheat,$^4$
T. Gamble,$^1$
L. K. Hardy,$^1$
R. D. H. Hickman,$^2$
P. Irawati,$^5$
\newauthor
D. J. Ives,$^6$
P. Kerry,$^1$
A. Leckngam,$^5$
S. P. Littlefair,$^1$
S. A. McLay,$^6$
K. O'Brien,$^7$
\newauthor
P. T. Peacocke,$^8$
S. Poshyachinda,$^5$
A. Richichi,$^5$
B. Soonthornthum,$^5$
A. Vick$^3$
\\
$^{1}$Department of Physics and Astronomy, University of Sheffield,
Sheffield S3 7RH, UK \\
$^{2}$Department of Physics, University of Warwick, Coventry CV4 7AL, UK \\
$^{3}$UK Astronomy Technology Centre, Royal Observatory Edinburgh,
Blackford Hill, Edinburgh EH9 3HJ, UK \\
$^{4}$Astrophysics Research Institute, Liverpool John Moores
University, Liverpool L3 5RF, UK \\
$^{5}$National Astronomical Research Institute of Thailand, 191
Siriphanich Building, Huay Kaew Road, Chiang Mai 50200, Thailand \\
$^{6}$European Southern Observatory, Karl-Schwarzschild-Str. 2,
D-85748 Garching bei M\"{u}nchen, Germany \\
$^{7}$Department of Physics, University of Oxford, Oxford OX1 3RH, UK \\ 
$^{8}$Lyncaeus Ltd., 1 George Street, Barnard Castle, Co. Durham DL12
8JD, UK}

\begin{document}

\date{Accepted 2014 August 12. Received 2014 August 12; in original form 2014 June 26.}

\pagerange{\pageref{firstpage}--\pageref{lastpage}} \pubyear{2014}

\maketitle

\label{firstpage}

\begin{abstract}
  ULTRASPEC is a high-speed imaging photometer mounted permanently at
  one of the Nasmyth focii of the 2.4-m Thai National Telescope (TNT)
  on Doi Inthanon, Thailand's highest mountain. ULTRASPEC employs a
  1024$\times$1024 pixel frame-transfer, electron-multiplying CCD
  (EMCCD) in conjunction with re-imaging optics to image a field of
  7.7'$\times$7.7' at (windowed) frame rates of up to
  $\sim$200\,Hz. The EMCCD has two outputs -- a normal output that
  provides a readout noise of 2.3\,e$^-$ and an avalanche output that
  can provide essentially zero readout noise. A six-position filter
  wheel enables narrow-band and broad-band imaging over the wavelength
  range 330--1000\,nm. The instrument saw first light on the TNT in
  November 2013 and will be used to study rapid variability in the
  Universe. In this paper we describe the scientific motivation behind
  ULTRASPEC, present an outline of its design and report on its
  measured performance on the TNT.
\end{abstract}


\section{Introduction}
\label{sec:introduction}

High-speed optical photometry can be defined as photometry obtained on
timescales of tens of seconds and below. This is a technological
definition, based on the fact that the conventional CCDs found on the
vast majority of the world's largest telescopes take tens of seconds
or longer to read out. Hence if one wishes to perform high-speed
optical photometry on large telescopes, it is usually necessary to
build specialised instruments dedicated to the task. A list of some of
the world's high-speed optical photometers is given by
\cite{dhillon07a}.

High-speed photometry enables the study of compact objects, such as
white dwarfs, neutron stars and black holes. This is because the
dynamical timescales of these stellar remnants range from seconds to
milliseconds, so that their rotation, pulsation, and the motion of any
material in close proximity to them (e.g. in an accretion disc), tends
to occur on these short timescales. Hence only by observing at high
speeds can the variability of compact objects be resolved, and in this
variability one finds encoded a wealth of information, such as their
structure, radii, masses and emission mechanisms,
e.g. \cite{littlefair06c}.

The study of white dwarfs, neutron stars and black holes, both
isolated and in binary systems, is of great importance in
astrophysics. For example, they allow us to test theories of
fundamental physics to their limits: black holes give us the chance to
study the effects of strong-field general relativity, and neutron
stars and white dwarfs enable the study of exotic states of matter
predicted by quantum mechanics. Black holes, neutron stars and white
dwarfs also provide a fossil record of stellar evolution, and the
evolution of such objects within binaries is responsible for some of
the Galaxy's most exotic and scientifically-valuable inhabitants, such
as accreting black holes, millisecond pulsars and binary white dwarfs,
e.g. \cite{gandhi10}, \cite{antoniadis13}.

The study of other classes of compact objects, with dynamical
timescales of minutes rather than seconds, can also benefit from
high-speed optical photometry. For example, one can use high
time-resolution observations of the eclipses, transits and
occultations of exoplanets, brown dwarfs and solar system objects to
provide unsurpassed spatial resolution, as well as to search for small
variations in their orbits due to the presence of other bodies,
e.g. \cite{zalucha07}, \cite{ortiz12}, \cite{marsh14},
\cite{richichi14}. Some of these targets, e.g. the host stars of
exoplanets, can also be very bright, and in these cases the short
exposures (with negligible dead time between exposures) are beneficial
to avoid saturation without destroying the efficiency of the
observations.

\section{A brief history of ULTRASPEC}
\label{sec:ultraspec}

In 2002, our team commissioned a high-speed, triple-beam imaging
photometer known as ULTRACAM \citep{dhillon07a}. Motivated by a desire
to understand the {\em kinematics} of compact objects, we then decided
to build a spectroscopic version of ULTRACAM, which we called
ULTRASPEC.  Two key factors drove the design of ULTRASPEC -- the first
was the requirement to minimise detector readout noise, as the light
in a spectrum is spread out over many more pixels than it is in an
image. This led us to use a frame-transfer electron-multiplying CCD
(EMCCD) -- see Sect.~\ref{sec:detector} -- as the detector in
ULTRASPEC. The second was the realisation that it was unnecessary to
build a new spectrograph for ULTRASPEC, as so many excellent
spectrographs with external focii able to accept visiting cryostats
already exist. Thus we designed ULTRASPEC as a bare (i.e. no optics)
EMCCD in a cryostat, to be bolted onto an existing spectrograph, and
re-using as much of the ULTRACAM data acquisition hardware and
software as possible, as described by \cite{ives08} and
\cite{dhillon08}. 

ULTRASPEC was tested for the first time on-sky during a 4-night
commissioning run in December 2006 on the EFOSC2 spectrograph mounted
on the ESO 3.6-m telescope at La Silla. This was the first time an
EMCCD had been used to perform astronomical spectroscopy on a
large-aperture telescope \citep{dhillon07b}, a topic which is
discussed in detail by \cite{tulloch11}. The first 17-night science
run with ULTRASPEC took place in January 2008 on the ESO 3.6-m, and a
second 22-night science run with EFOSC2 mounted on the 3.5-m New
Technology Telescope (NTT) at La Silla was successfully completed in
June 2009.

Visiting instruments such as ULTRACAM and ULTRASPEC have the major
disadvantage that they are subject to the vagaries of the
time-allocation process and are mounted on a telescope for only a few
weeks each year, making it difficult to respond rapidly to new
astronomical discoveries and to plan long-term monitoring or survey
observations. So in 2010 our priorities switched from performing
high-speed spectroscopy with ULTRASPEC to finding a permanent home for
the instrument and converting it into a high-speed photometer. In
2011, we signed a Memorandum of Understanding with the National
Astronomical Research Institute of Thailand (NARIT), providing the
ULTRASPEC team with 30 nights per year on the new 2.4-m Thai National
Telescope (TNT) in return for access to ULTRASPEC for the Thai
astronomical community for the rest of the year. In the course of the
move to the TNT, ULTRASPEC's data acquisition hardware and software
were upgraded to bring them into line with the latest version of the
ULTRACAM data acquisition system. We also constructed an
opto-mechanical chassis on which to mount the original ULTRASPEC
cryostat and a new set of re-imaging optics. ULTRASPEC saw first light
on the TNT on 2013 November 5.

No detailed description of ULTRASPEC, in either its spectroscopic or
photometric incarnations, has appeared in the refereed astronomical
literature. In this paper, therefore, we describe the design and
performance of ULTRASPEC, primarily in its new role as a high-speed
optical photometer mounted on the TNT.

\section{Design}
\label{sec:design}

We begin this section with a brief summary of the TNT, followed by a
detailed description of the design of ULTRASPEC in its current
form as a high-speed imaging photometer on the TNT.

\subsection{The Thai National Telescope}
\label{sec:tnt}

\begin{figure*}
\centering
\includegraphics[width=8.5cm,angle=0]{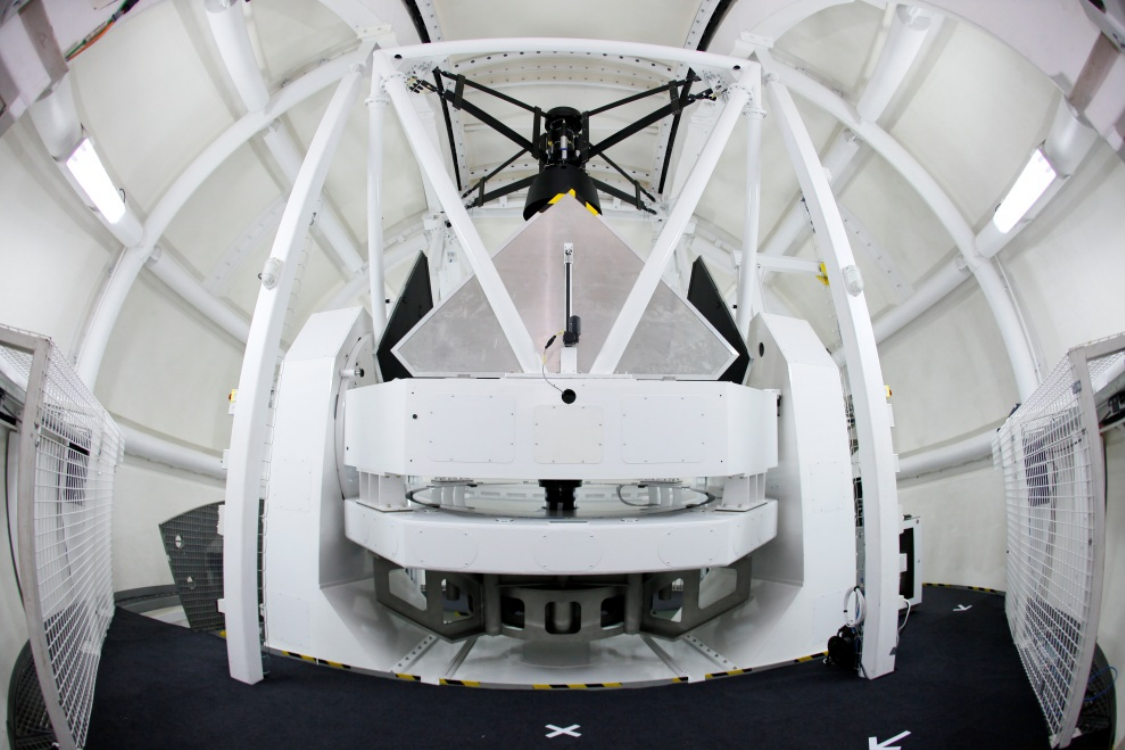}\hspace{0.5cm}\includegraphics[width=8.5cm]{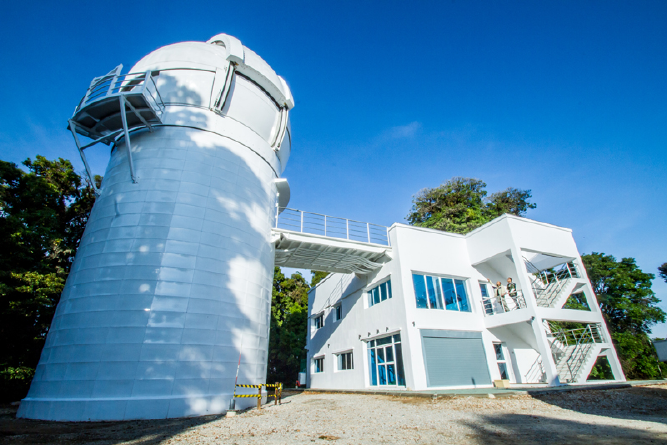}
\caption{Left: Interior photograph of the TNT. The mirror covers are
  open and the 2.4-m $f$/1.5 primary mirror can be seen towards the
  bottom. Note the extremely compact telescope design and the
  innovative dome composed of a tubular-steel support structure
  covered in fibre-glass panels. Right: Exterior photograph of the
  TNT, showing the tower on which the telescope sits and the adjacent
  control-room building.}
\label{fig:tnt}
\end{figure*}

The Thai National Telescope is a 2.4-m Ritchey-Chr\'{e}tien with an
$f$/10 focal ratio, providing a plate scale of 8.6''/mm at the two
Nasmyth focii (Fig.~\ref{fig:tnt}). The telescope is on an alt-az
mount which can slew at up to 4$^{\circ}$/s, making it ideal for the
rapid acquisition of transient sources triggered by other
facilities. The pointing accuracy is better than 3'' over most of the
sky and the telescope can track to better than 0.5'' over 10\,min
without autoguiding. The TNT was designed and built by EOS
Technologies, USA/Australia, and saw first light in early 2013. The
telescope is housed in a hemispherical fibre-glass dome, designed and
manufactured in Thailand.

The TNT (latitude 18.573725$^{\circ}$N, longitude
98.482194$^{\circ}$E, altitude 2457\,m) is the main facility of the
Thai National Observatory (TNO), located close to the summit of the
highest mountain in Thailand, Doi Inthanon. The site is approximately
100\,km by road from the nearest city -- Chiang Mai -- and lies in the
middle of a national park, thereby ensuring dark skies: $B = 22.5$ and
$V = 21.9$ magnitudes per square arcsecond on a moonless night,
comparable with the best observing sites in the world. Doi Inthanon is
tree-covered all the way to the summit, and the trees are protected
from being felled. Hence the telescope had to be built on a tower to
provide a clear view to the horizon (Fig.~\ref{fig:tnt}). The median
seeing is approximately 0.9'' and is remarkably stable on most nights,
rarely exceeding 2''. The climate at Doi Inthanon is split into wet
and dry seasons. The telescope is only operated during the dry season,
which runs from the beginning of November to the end of April. The
best months during the 2013/14 observing season were December--March,
with approximately 75\% usable nights during this period, similar to
that observed at other major observatories, e.g. La Palma
\citep{dellavalle10}.

At present, only one of the two Nasmyth focii is in operation. This
focus is equipped with a rotator, on which is mounted a cube with four
ports for instruments. Light from the tertiary (M3) mirror is directed
to a quarternary (M4) mirror mounted inside the cube, which can then
direct the beam to whichever of the four ports the astronomer wants to
use. At present, the four ports currently harbour: ULTRASPEC, an
autoguider, a fibre-fed spectrograph, and a conventional
4\,k$\times$4\,k CCD camera.

\subsection{Optics}
\label{sec:optics}

\begin{figure*}
\centering
\includegraphics[width=14cm,angle=0]{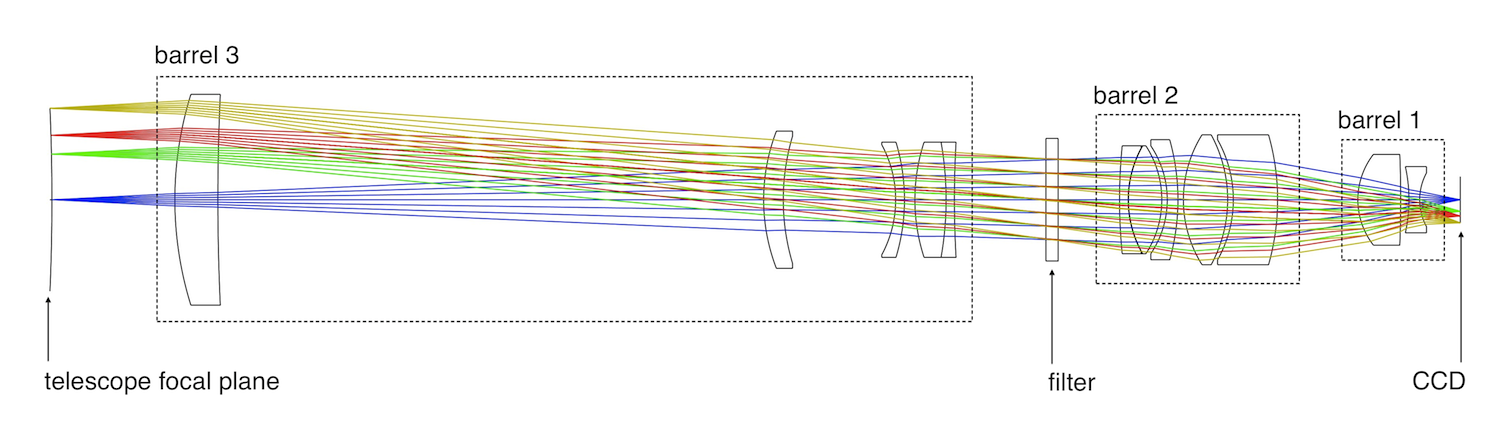}
\includegraphics[width=14cm]{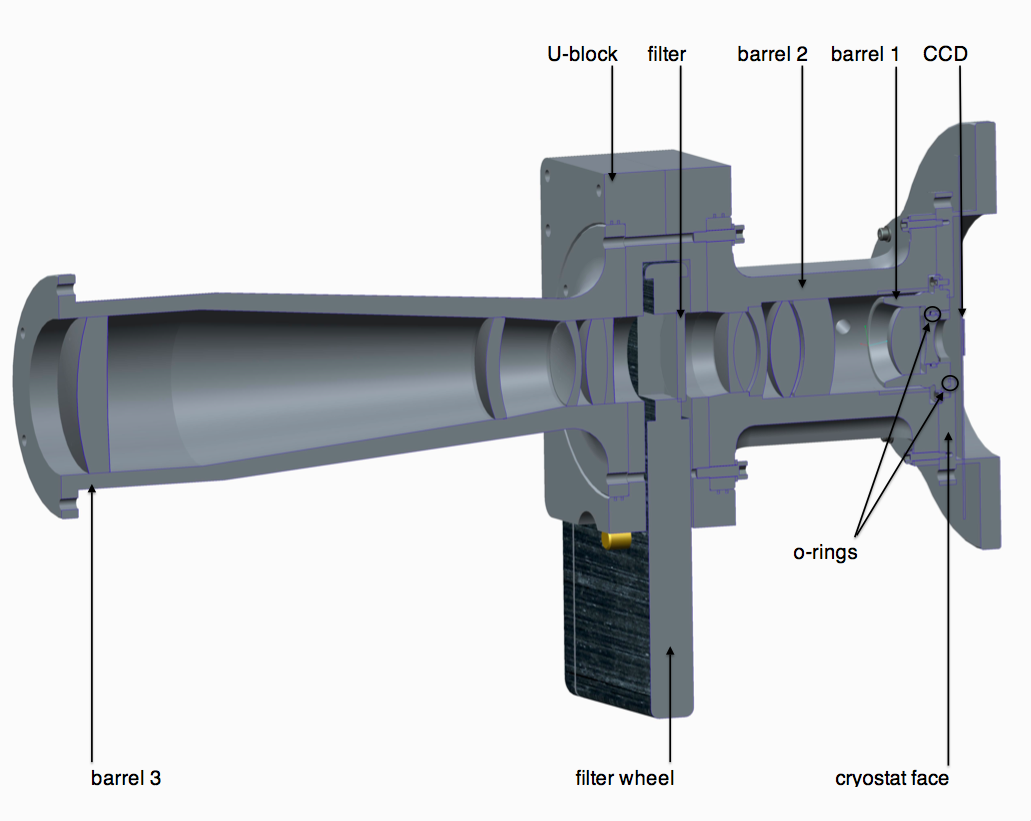}
\caption{Top: Ray-trace through the ULTRASPEC optics, showing the 12
  lenses and how they have been arranged into the three lens barrels.
  The telescope focal plane, filter and CCD surface are also
  indicated.  The diagram is to scale: the largest lens is in barrel~3
  and has a diameter of 86\,mm; the separation between the telescope
  focal plane and the CCD is 570\,mm. Bottom: Cross-section through
  the optics barrels, highlighting some of the components described in
  the text.}
\label{fig:raytrace}
\end{figure*}

Critically sampling the median seeing at the TNT requires a platescale
of 0.45''/pixel, which would result in a field of view of
7.7'$\times$7.7' on ULTRASPEC's 1024$\times$1024-pixel detector. Using
the star counts listed by \cite{simons95}, such a field of view would
result in an 80\%\ probability of finding a comparison star of
magnitude $R=11$ at a galactic latitude of 30$^{\circ}$ (the all-sky
average). This is brighter than the vast majority of our target stars,
thereby ensuring reliable differential photometry. Hence we designed a
4$\times$ focal reducer for ULTRASPEC in order to provide a platescale
of 0.45''/pixel, which can work across the desired wavelength range of
330--1000\,nm covered by the SDSS filters (see below).

\begin{table}
\centering

\caption{ULTRASPEC optical prescription. Note that this is the optical
  design after fitting to the manufacturer's test plates, and is not
  the as-built design incorporating the melt data and post-manufacture
  optimisation of the lens spacings. Lenses 4 and 5 are doublets. All
  dimensions are in mm. ROC = radius of curvature; Thickness = lens
  thickness or separation between surfaces; TFP = telescope focal
  plane. Two different anti-reflection coatings were used, marked 
  with superscripts 1 and 2, giving an average transmission 
  greater than 98\%\ across the 330-1000\,nm range on each 
  surface -- see Fig.~\ref{fig:filters}.}

\begin{tabular}{lrrl}
\hline\noalign{\smallskip}
Surface & \multicolumn{1}{c}{ROC} & \multicolumn{1}{c}{Thickness} & Glass \\
\hline\noalign{\smallskip}
TFP & & 50.046 & \\
Lens 1 & 144.930 & 17.970 & N-FK51A$^2$ \\
& 931.800 & 220.924 & \\
Lens 2 & 77.985 & 8.000 & N-BK7$^2$ \\
& 90.995 & 43.511 & \\
Lens 3 & --47.345 & 3.730 & N-PSK53A$^2$ \\
& --81.505 & 2.351 & \\
Lens 4 & 75.800 & 12.740 & N-BK7$^2$ \\
&  --134.940 & 3.500 & N-LAK21$^1$ \\
&  412.675 & 38.684 & \\
Filter & Infinity & 5.000 & SILICA \\
& Infinity & 25.000 & \\
Lens 5 & 223.375 & 3.500 & N-SK14$^1$ \\
& 44.605 & 12.570 & CAF2$^2$ \\
& --40.600 & 2.194 & \\
Lens 6 & --39.460 & 4.460 & N-LAK21$^1$ \\
& --90.995 & 1.911 & \\
Lens 7 & 50.065 & 18.000 & CAF2$^2$ \\
& --55.690 & 2.315 & \\
Lens 8 & --50.065 & 18.000 & N-BK7$^2$ \\
& --96.860 & 33.544 & \\
Lens 9 & 30.255 & 17.450 & N-BK7$^2$ \\
& --475.220 & 3.358 & \\
Lens 10 & --62.415 & 4.510 & N-LAK21$^1$ \\
& 23.070 & 16.244 & \\
CCD & Infinity & & \\
\hline\noalign{\smallskip}
\end{tabular}
\label{tab:optdes}
\end{table}

The final ULTRASPEC optical design is shown in Fig.~\ref{fig:raytrace}
and Tab.~\ref{tab:optdes}, and is composed of 12 lenses plus the
plane-parallel filter. The first 5 elements form a collimator that
images the telescope exit pupil onto the filters in the wheel.  These
lenses are packaged into barrel~3. The next 7 lenses form a camera
section, with sufficient space between camera and collimator to
accommodate the filter wheel assembly.  These lenses are packaged into
lens barrels~1~and~2. Since the pupil image is remote from both
collimator and camera, the collimator needs to act rather like an
eyepiece, and a modified Petzval type lens with field flattener was
chosen as the basis of the camera design. The main difficulty of the
design was presented by the very broad spectral range and the need for
co-registration of image centroids across all bands (correction of
lateral colour), which is essential if we want to use small windows
and change filter without having to adjust the telescope position
and/or window parameters each time.  In particular, the availability
of glasses with reasonable transmission over the
scientifically-important $u'$ band restricted the usable glass
types. The seeing conditions at the telescope site and the pixel size
determined the target modulation transfer function and spot size for
optimisation, with the goal of providing seeing-limited stellar images
in median seeing conditions (0.9'') at the TNT. A simple ghost
analysis finds the best-focused ghost to have an RMS spot area of 47
times the image spot area, with a brightness 0.0002\%\ of the stellar
image and a displacement that varies from zero on axis to 145\,$\mu$m
($\sim11$ pixels) at the edge of the field.

There are two doublets in the optical design, lenses 4 and 5 in
Tab.~\ref{tab:optdes}. N-BK7 and N-LAK21 are both non-crystalline
glasses with similar coefficients of thermal expansion. Hence we were
able to use a relatively rigid cement with excellent transmission
properties -- Norland Optical Adhesive NOA88. N-SK14 and CaF$_2$ have
different thermal expansion properties, resulting in a radial
expansion of 8.3\,$\mu$m when the temperature fluctuates by 30\,K
around room temperature, giving significant shear stress on any
adhesive used. This is particularly problematic for the CaF$_2$
element which is crystalline and less resilient. Hence we chose to
bond these two elements using a 120\,$\mu$m layer of flexible
silicon-based adhesive known as RTV\,615. The lenses were manufactured
by IC Optical Systems, UK, and coated by CVI Melles Griot, Isle of
Man. The lens barrels were manufactured, and the lenses mounted and
aligned within them, by IC Optical Systems.

\begin{figure}
\centering
\hspace*{-0.1cm}\includegraphics[width=8.5cm,angle=0]{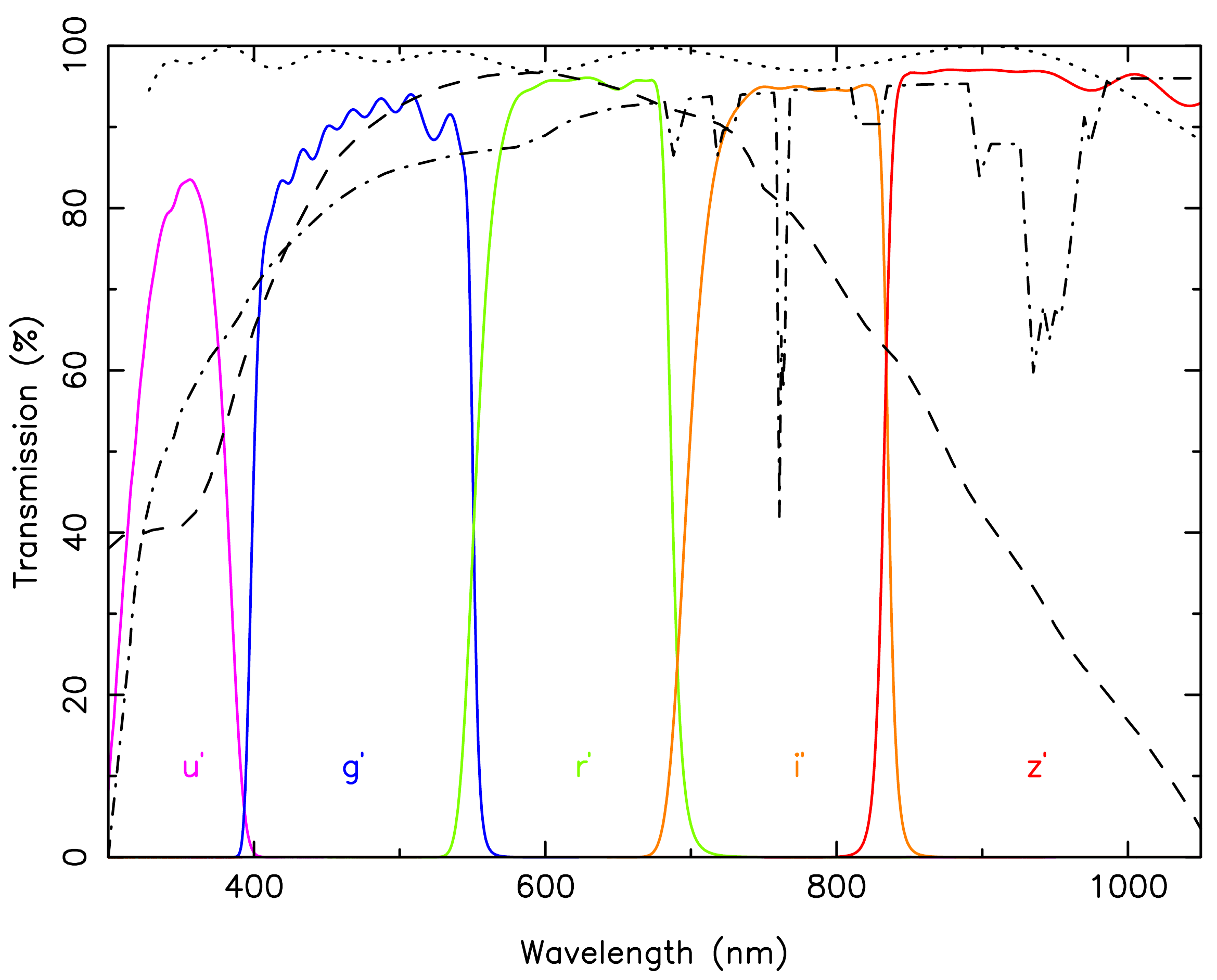}
\caption{Transmission profiles of the ULTRASPEC SDSS filter set
  (purple, blue, green, orange and red solid lines correspond to $u'$,
  $g'$, $r'$, $i'$ and $z'$, respectively). The transmission of one of
  the anti-reflection coatings used on the ULTRASPEC lenses (the
  CaF$_2$ elements, dotted line), and the transmission of the
  atmosphere for unit airmass (dashed-dotted line), are also
  shown. The quantum efficiency curve of the ULTRASPEC EMCCD, which
  uses e2v's standard midband anti-reflection coating, is shown by the
  dashed line.}
\label{fig:filters}
\end{figure}

The Sloan Digital Sky Survey (SDSS) photometric system
\citep{fukugita96} was adopted as the primary filter set for ULTRASPEC
-- the $u'$\,$g'$\,$r'$\,$i'$\,$z'$ pass-bands are shown in
Fig.~\ref{fig:filters}. We chose this filter set primarily because we
wanted to ensure that we could combine our ULTRASPEC observations with
those made with our other high-speed imaging photometer, ULTRACAM
\citep{dhillon07a}. As well as SDSS filters, which are identical to
those used in ULTRACAM, ULTRASPEC also has an extensive set of
broad-band and narrow-band filters -- see Tab.~\ref{tab:filters}.
Only 6 filters can be mounted at any one time in the wheel. All
ULTRASPEC filters are 50$\times$50 mm$^2$ and approximately 5 mm
thick, but have been designed to have identical {\em optical}
thicknesses so that their differing refractive indices are compensated
by slightly different physical thicknesses, making the filters
interchangeable without having to refocus the telescope. The filters
have also been designed taking the ULTRASPEC optical design into
account to ensure that the required central wavelength and FHWM are
achieved. All of the filters used in ULTRASPEC were designed and
manufactured by Asahi Spectra Company, Japan.

\begin{table}
\centering
\caption{ULTRASPEC filters. $\lambda_{c}$ is the central wavelength
  and $\Delta\lambda$ is the FWHM. The clear filter is approximately
  $u' + g' + r' + i' + z'$ and the Schott KG5 filter is approximately
  $u' + g' + r'$.}
\begin{tabular}{lcc}
\hline\noalign{\smallskip}
Filter & $\lambda_{c}$ (nm) & $\Delta\lambda$ (nm) \\
\hline\noalign{\smallskip}
$u'$ & 355.7 & 59.9 \\
$g'$ & 482.5 & 137.9 \\
$r'$ & 626.1 & 138.2 \\
$i'$ & 767.2 & 153.5 \\
$z'$ & 909.7 & 137.0 \\
Clear & -- & -- \\
Schott KG5 & 507.5 & 360.5 \\
$i' + z'$ & 838.5 & 290.5 \\
CIII/NIII+HeII & 465.7 & 11.2 \\
Blue continuum & 514.9 & 15.8 \\
NaI & 591.1 & 31.2 \\
Red continuum & 601.0 & 11.8 \\
H$\alpha$ wide & 655.4 & 9.4 \\ 
H$\alpha$ narrow & 656.4 & 5.4 \\
\hline\noalign{\smallskip}
\end{tabular}
\label{tab:filters}
\end{table}

\subsection{Mechanics}
\label{sec:mechanics}

\begin{figure*}
\centering
\includegraphics[width=10cm,angle=0]{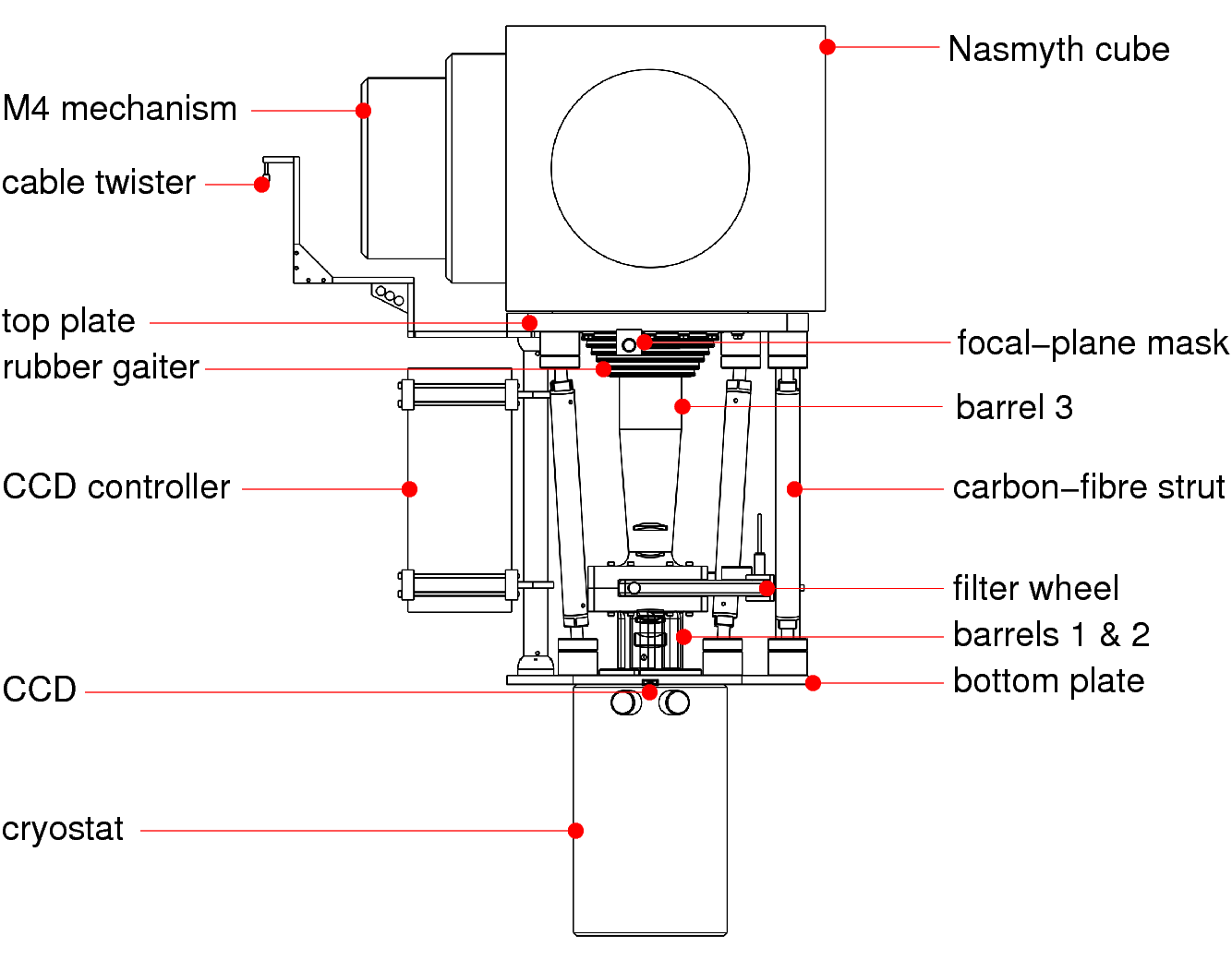}\hspace{0.5cm}\includegraphics[width=7cm]{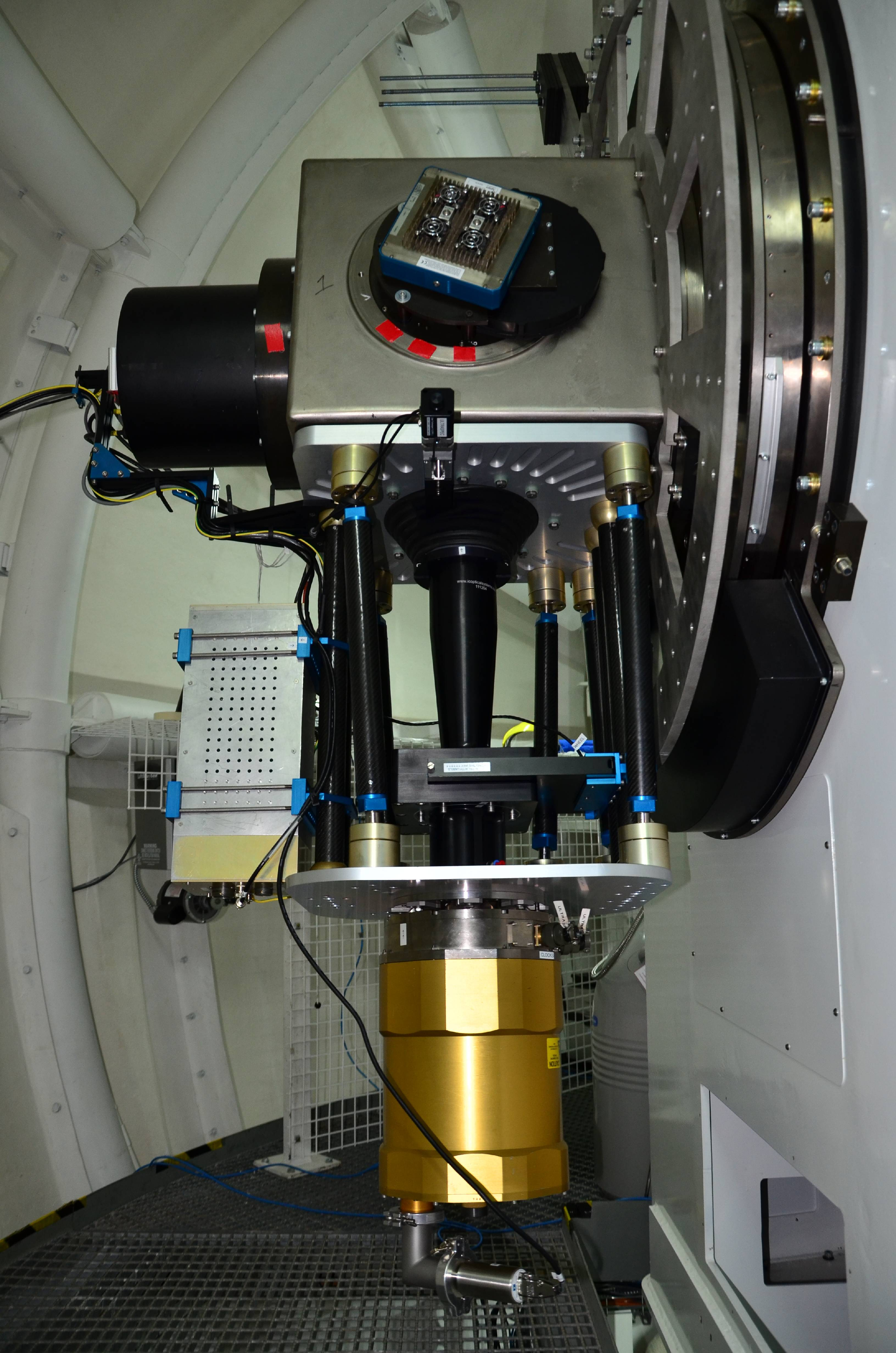}
\caption{Left: Mechanical drawing of the ULTRASPEC opto-mechanical
  chassis, highlighting some of the components shown in the photograph
  on the right and described in the text. Right: Photograph of
  ULTRASPEC mounted on the Nasmyth focus of the TNT. The length and
  mass of ULTRASPEC, not including the Nasmyth cube, are
  approximately 1.0\,m and 90\,kg.}
\label{fig:mech}
\end{figure*}

The design of the mechanical structure of ULTRASPEC on the TNT was
driven by the requirement to fit within the space envelope and mass
limit determined by the properties of the Nasmyth rotator and
cube. Specifically, any instrument mounting on the cube has to be less
than 1\,m in length and weigh no more than 100\,kg. This proved quite
challenging, given the dimensions and masses of the optics barrels and
the cryostat, so we decided upon an open-strut design, utilising
aluminium and carbon fibre wherever possible for their low mass and
high rigidity. The other requirements for our opto-mechanical chassis
were: i) to provide a platform on which to mount the optics, cryostat
and CCD controller; ii) to allow easy access to the optics and
cryostat for filter changes and maintenance; iii) to exhibit low
thermal expansion to minimise focus changes with temperature; iv) to
exhibit low flexure (of order 1 pixel, i.e. 13 $\mu$m) at any
orientation, so stars do not drift out of the small windows defined on
the CCD; v) to be electrically isolated from the telescope in order to
reduce pickup noise via ground loops.

The mechanical structure of ULTRASPEC, which meets all of the
requirements described above, is shown in Fig.~\ref{fig:mech}. Two
aluminium plates are held parallel with respect to each other via a
set of 10 carbon-fibre struts. The top plate attaches to the
Nasmyth cube and the bottom plate has the cryostat attached
to it. To ensure that the three optics barrels do not take the weight
of the instrument, they are mounted on the front face of the cryostat
(see also Fig.~\ref{fig:raytrace}) and do not touch any other part of
the instrument -- a rubber gaiter between the optics and top plate
provides a light and dust-proof seal. This way, any flexure in the
mechanical structure does not cause corresponding flexure in the
optics barrels, minimising any degradation in image quality. In order
to minimise the mass of the optics train, and hence minimise flexure
within it, the filter wheel is mounted on the carbon-fibre
struts. The filter wheel sits within the `U-block' that mechanically
couples lens barrels 2 and 3 (Fig.~\ref{fig:raytrace}), but does not
touch it. The CCD controller, which must reside as close as possible
to the cryostat in order to minimise cable lengths and hence pickup
noise, is also mounted on the carbon-fibre struts.

In order to maximise throughput, we decided to use an active cryostat
window, i.e. instead of using a plane-parallel window, the right-most
lens shown in Fig.~\ref{fig:raytrace} acted as the cryostat
window. The difficulty with this design is that the lens then has to
form a vacuum seal, and it must do this in such a way that its
position with respect to the other lenses and the CCD does not change
as the pressure in the cryostat changes, e.g. due to the changing
compression of an o-ring. We met this requirement by polishing the rim
of the lens, and then placing an o-ring between the rim of the lens
and the inner wall of the aluminium lens barrel, as shown in
Fig.~\ref{fig:raytrace}. A second o-ring was then used to seal the gap
between the lens barrel itself and the front face of the cryostat,
with suitable slots machined in the barrel to eliminate trapped
volumes, which would have compromised the vacuum. Note that the amount
of compression of this second o-ring was fixed to a known value by
using a suitably machined slot for the o-ring.  The position of the
CCD within the cryostat was measured using a travelling microscope and
adjusted using shims so that it is flat and at the correct depth (when
cold) with respect to the optics, to an accuracy of better than
0.1\,mm.

For ease of cryostat maintenance, we kept lens barrel 1
small, containing only the two lenses closest to the CCD. Lens barrel
1 fits inside lens barrel 2, which is much more bulky as it is
designed to take the weight of the U-block and lens barrel 3. To
prevent any condensation on the outer lens of barrel 1 in humid
conditions, we blow dry nitrogen gas into the cavity between lens
barrels 1 and 2.

A layer of G10/40 isolation material is placed between the cube and
the top plate of ULTRASPEC to provide electrical isolation from the
telescope. The top plate also has a motorised focal-plane mask mounted
on it. This is an aluminium blade that can be moved into the focal plane
to prevent light from falling on regions of the CCD chip outside the
user-defined windows typically used for observing. Without this mask,
light from bright stars falling on the active area of the chip above
the CCD windows would cause vertical streaks in the windows -- see
Fig.~1 of \cite{dhillon05} for an example. The mask also prevents
photons from the sky from contaminating the background in drift-mode
windows (see Sect.~\ref{sec:modes}). The mechanical structure
incorporates a cable twister, composed of an aluminium arm along which
all of the ULTRASPEC cables are routed to a point lying on the
rotation axis of the instrument. The cables are clamped at this point,
and then run horizontally for approximately 1\,m to another attachment
point on the dome, which is slaved to the telescope, where they are
also clamped. This freely-suspended 1\,m cable bundle is then able to
accommodate any twisting imposed by the Nasmyth rotator.

\begin{figure*}
\centering
\includegraphics[width=11cm]{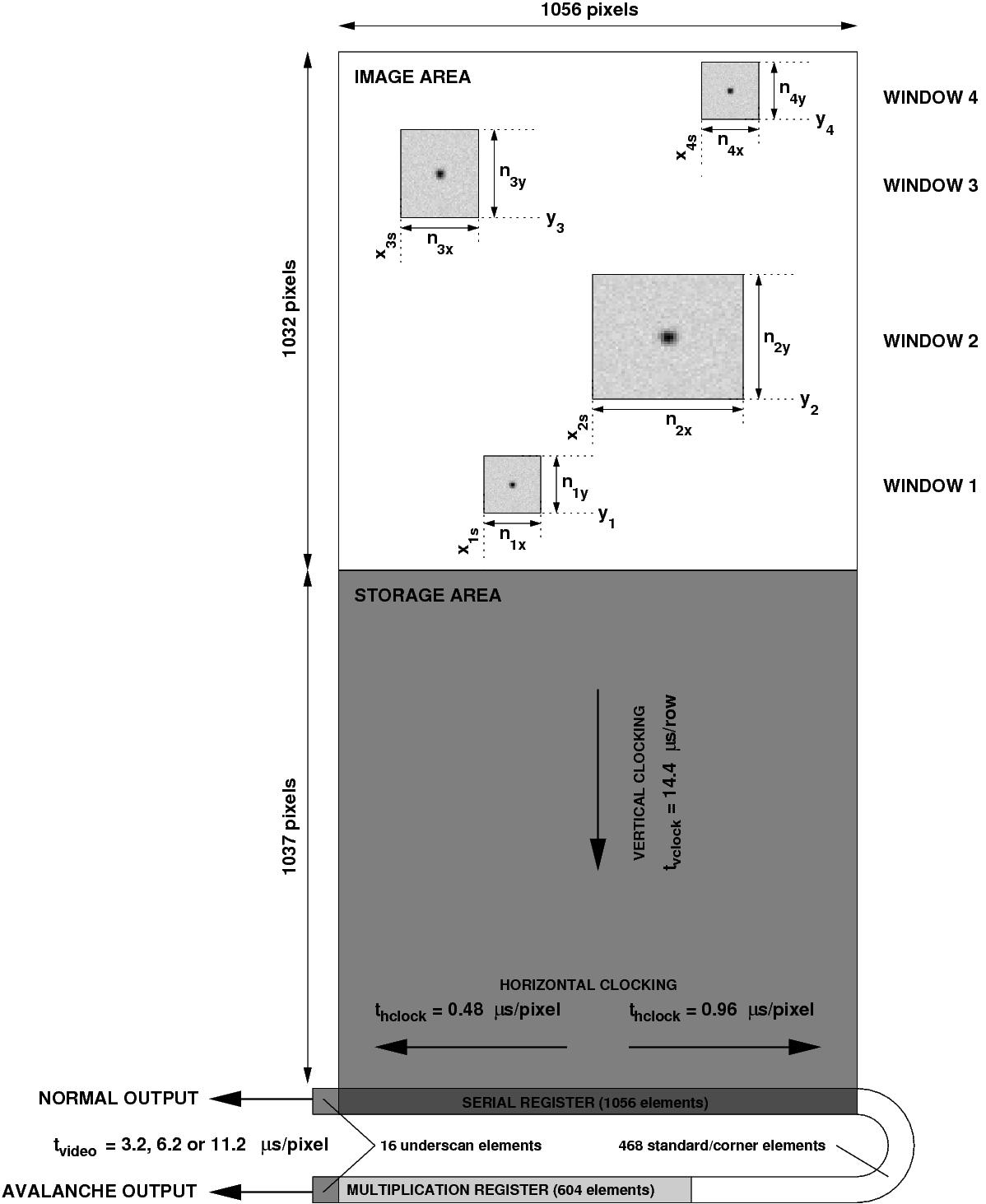}\hspace{0.5cm}\includegraphics[width=6cm]{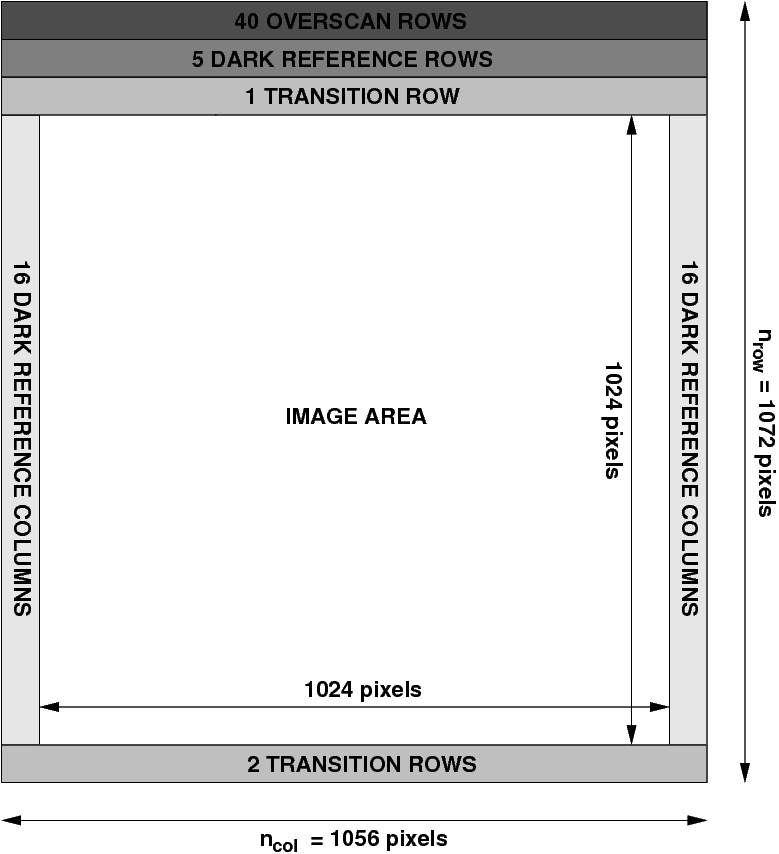}
\vspace*{0.1cm}
\caption {Left: Pictorial representation of the ULTRASPEC EMCCD. The
  pixel-dimensions and locations of the image area, storage area,
  serial register and multiplication register are shown, along with
  the vertical and horizontal clocking directions. It can be seen that
  there are two outputs: clocking to the left sends the electrons to
  the normal output, whilst clocking to the right sends them to the
  avalanche output. Up to 4 windows can be defined, or the chip can be
  read out in full-frame mode, in which case the entire image area is
  selected (see Fig.~\protect\ref{fig:first_light}). Note that the
  16 serial underscan elements just prior to each output are skipped
  over and do not form part of the final output image shown in the
  right-hand panel. The various window parameters and pixel rates are
  also shown. Right: Schematic showing the format of an ULTRASPEC data
  frame, as output by the data reduction pipeline
  (\protect{Sect.~\ref{sec:pipeline}}), which is of size
  1056$\times$1072 pixels. The diagram shows the locations of the
  various transition rows, dark reference rows/columns and overscan
  elements. The latter are not real pixels and is the only region that
  should be used for bias-level determination. Only the central
  1024$\times$1024 pixels marked on the diagram are sensitive to
  light.}
\label{fig:frame}
\end{figure*}

\subsection{Detector}
\label{sec:detector}

ULTRASPEC uses an e2v CCD\,201-20 as its detector. This is a
frame-transfer EMCCD with an image area of 1024$\times$1024 pixels,
each of size 13 $\mu$m -- see Fig.~\ref{fig:frame}. The chip is
thinned, back-illuminated and coated with e2v's standard midband
anti-reflection coating, providing a maximum quantum efficiency (QE)
of approximately 96\%\ around 600\,nm -- see Fig.~\ref{fig:filters}.
The device used in ULTRASPEC is a grade 1 device, i.e. it is of the
highest cosmetic quality available. The CCD\,201-20 is a two-phase
device, thereby minimising the number of vertical clocks required and
thus maximising the frame rate. The full well capacity of each pixel
is approximately 80\,000\,e$^-$, and we operate it with a system gain
of $g_{S}\sim0.8$\,e$^-$/ADU to match the 16-bit analogue-to-digital
(ADC) converter in the CCD controller (see Sect.~\ref{sec:das}). The
CCD has been measured to be linear to within 1\% up to the saturation
level of the ADC.

Although it is possible to run the CCD\,201-20 in inverted mode to
suppress dark current at higher operating temperatures, this has been
shown to increase the rate of clock-induced charge (CIC; see
\citealt{tulloch11}). As discussed by these authors, CIC becomes a
significant source of noise when operating EMCCDs in
electron-multiplying mode (see below), and so we operate the ULTRASPEC
EMCCD in non-inverted mode around 160\,K. This results in negligible
dark current of 10\,e$^{-}$/pixel/hr and a clock-induced charge rate
of approximately 0.02 e$^{-}$/pixel/frame. To reduce conductive and
convective heating, and to prevent condensation on the CCD, the
cryostat is evacuated to a pressure of approximately $10^{-6}$\,mbar,
which it is able to maintain for a period of at least 6 months without
further pumping. The cryostat is cooled with liquid nitrogen, and the
temperature of the CCD is regulated with a Lakeshore 331 temperature
controller. The volume of the cryostat is 2.5\,litres, giving a hold
time of approximately 20 hours.

Fig.~\ref{fig:frame} shows that the CCD\,201-20 has two outputs,
which can be easily switched in software. Sending the photo-electrons
to the normal output allows the CCD to operate in an identical manner
to a conventional CCD, with a readout noise of 2.3\,e$^{-}$ at a pixel
rate of 89\,kHz. This is the preferred option when the signal-to-noise
ratio (SNR) of an observation is limited by shot noise from the object
or sky. Sending the electrons to the avalanche output, on the other
hand, results in on-chip amplification of the electrons in the
multiplication register by a factor of $g_{A} \sim 1000$, via a
process known as impact ionisation. The resulting readout noise is
reduced by a factor $1/g_{A}$, rendering it negligible. This is the
preferred option when the SNR of an observation is readout-noise
limited, although it should be noted that the optimum choice of output
is not quite as straight-forward as this due to the stochastic nature
of the electron multiplication process, which effectively reduces the
QE by a factor of 2 when not photon counting. We discuss the relative
merits of the normal and avalanche outputs further in
Sect.~\ref{sec:sensitivity} and refer the reader to \cite{tulloch11}
and references therein for a more in-depth discussion of EMCCDs.

The various clocking and pixel rates used in ULTRASPEC are indicated
in Fig.~\ref{fig:frame}. The vertical clocking rate is
14.4\,$\mu$s/row and the storage area contains 1037 rows, resulting in
a frame-transfer time of 14.9\,ms. This is the dead time
between exposures, unless drift mode is being used (see
Sect.~\ref{sec:modes}). Three readout speeds are currently available
with ULTRASPEC: slow (11.2~$\mu$s/pixel), medium (6.2 $\mu$s/pixel)
and fast (3.2 $\mu$s/pixel), giving full-frame cycle times of 12.8,
7.2 and 3.8\,s and readout noise of 2.3, 2.7 and 4.4\,e$^{-}$,
respectively, from the normal output.

\subsection{Data acquisition system}
\label{sec:das}

The key requirement for the data acquisition system for a high-speed
imaging photometer is that it must be {\em detector limited}, i.e. the
throughput of data from the output of the CCD to the hard disk on
which it is eventually archived is always greater than the rate at
which the data comes off the CCD. The ULTRASPEC data acquisition
system meets this requirement, and hence is capable of running
continuously all night without having to pause for archiving of
data. It is very similar to the data acquisition system of ULTRACAM,
differing primarily in the requirement to operate a single EMCCD as
opposed to three conventional frame-transfer CCDs. In this section we
provide a brief overview of the ULTRASPEC data acquisition system -- a
much more detailed description can be found in \cite{beard02},
\cite{ives08} and \cite{mclay10}.

\subsubsection{Hardware}

\begin{figure}
\centering
\includegraphics[width=8.5cm]{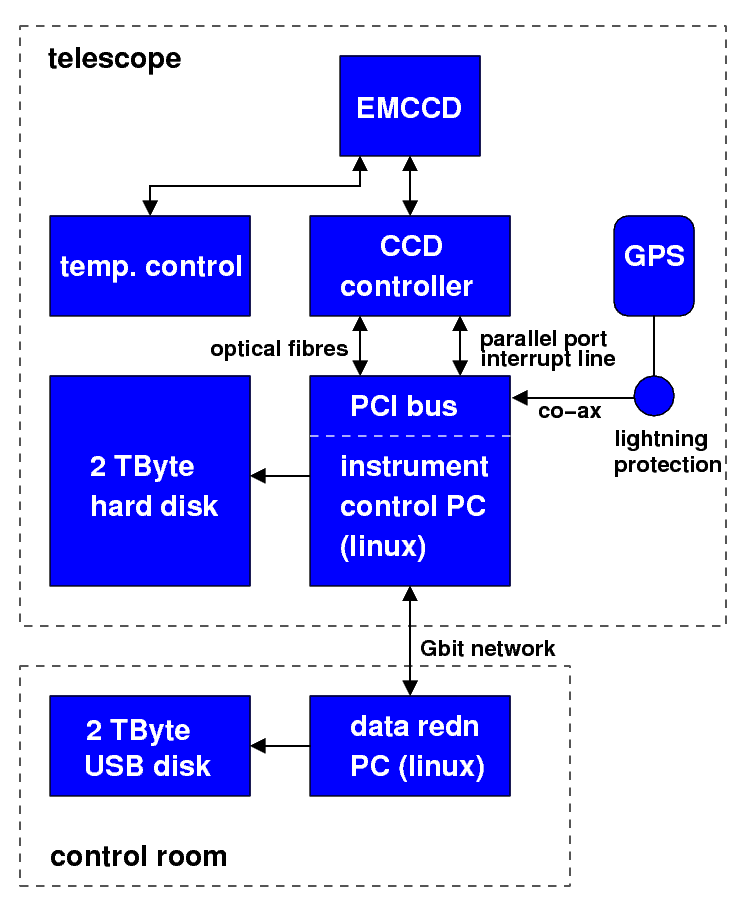}
\caption{Schematic showing the principal hardware components of the
ULTRASPEC data acquisition system.  The connections between the
hardware components and their locations at the telescope are also
indicated.}
\label{fig:das}
\end{figure}

Fig.~\ref{fig:das} shows the data acquisition hardware used in
ULTRASPEC. Data from the EMCCD is read out by an Astronomical Research
Cameras (ARC) Generation III CCD controller, also known as a San Diego
State University (SDSU) controller \citep{leach00}. The SDSU
controller is hosted by a rack-mounted, quad-core PC running Linux
patched with RealTime Application Interface (RTAI) extensions. RTAI is
used to provide strict control over one of the processor cores, so as
to obtain accurate timestamps from the Global Positioning System (GPS)
antenna connected to the PC (see Sect.~\ref{sec:gps} for
details). Note that the GPS antenna is located inside the TNT dome, as
the fibre-glass skin is transparent to GPS frequencies.

The instrument control PC communicates with the SDSU controller via a
Peripheral Component Interconnect (PCI) card and two 250 MHz optical
fibres. As well as communicating through the fibres, the SDSU
controller has the ability to interrupt the PC using its parallel port
interrupt line, which is required to perform accurate timestamping
(see Sect.~\ref{sec:gps}). Data from the CCD is passed from the
SDSU PCI card to the PC memory by Direct Memory Access (DMA), from
where the data is written to a high-capacity hard disk. All of the
work in reading out the CCD is performed by the SDSU controller; the
PCI card merely forwards commands and data between the instrument
control PC and the SDSU controller.

To enable electron-multiplication operation, we designed a
high-voltage clock board for the SDSU controller. This provides the
40\,V clock swings required for impact ionisation to occur in the
multiplication register. The voltage output by the board is very
stable and user-adjustable in 9 steps, providing
electron-multiplication factors of up to $\sim 1000$ (see
\cite{ives08} for details). In practice, unless we are worried about
saturating the multiplication register (see
Sect.~\ref{sec:sensitivity}), we only use the highest
electron-multiplication setting as this gives the lowest effective
readout noise.

\subsubsection{Software}

The SDSU controller and PCI card both have on-board digital signal
processors (DSPs) which can be programmed by downloading assembler
code from the instrument control PC. An ULTRASPEC user wishing to take
a sequence of windowed images, for example, would load the relevant
DSP application onto the SDSU controller (to control the CCD) and PCI
card (to handle the data). The user can also modify certain
parameters, such as the exposure time, readout speed, window
sizes/positions and binning factors, by writing the new values
directly to the DSP's memory.

All communication within the ULTRASPEC data acquisition system,
including the loading of DSP applications on the SDSU controller, is
via Extensible Markup Language (XML) documents transmitted using the
Hyper Text Transfer Protocol (HTTP) protocol. This is an international
communications standard, making the ULTRASPEC data acquisition system
highly portable and enabling users to operate the instrument using any
interface able to send XML documents via HTTP, e.g. web browser,
Python script, Java graphical user interface (GUI). In the latest
version of the ULTRASPEC data acquisition system, a Python GUI runs on
the Linux data reduction PC in the control room. This allows the
astronomer to set the various CCD parameters, and the GUI then
automatically writes and sends an XML document to execute the
observations.

\subsection{Timestamping}
\label{sec:gps}

ULTRASPEC can image at rates of up to $\sim$200\,Hz, hence it is critical
that each frame is timestamped to an accuracy of a millisecond or
better. Whenever a new exposure is started, as denoted by a
frame-transfer operation, the SDSU controller sends an interrupt, in
the form of a voltage step sent down a co-axial cable, to the
instrument control PC. Owing to the use of RTAI, the PC then {\em
  instantaneously} (i.e. within $\sim 10\,\mu$s) reads the current
time from a GPS-synchronised clock on board a Meinberg GPS170PEX PCI
Express card and writes the current time to a First-In First-Out
(FIFO) buffer. The data handling software then reads the timestamp
from the FIFO and writes it to the header of the next buffer of raw
data written to the PC memory. In this way, the timestamps and raw
data always remain synchronised.  

Theoretically, the absolute accuracy of the ULTRASPEC timestamping
should be of order tens of microseconds. In order to measure this, we
observed with ULTRASPEC an LED triggered by the pulse-per-second (PPS)
output of the Meinberg GPS card. The PPS output is accurate to better
than 250\,ns and the rise time of the LED is of a similar order, hence
these are both negligible sources of error. We measured an LED turn-on
time of $0.000\pm0.001$\,s in both drift mode and one-window mode,
verifying that the absolute timestamping accuracy of ULTRASPEC is
better than 1\,ms. This test is obviously insensitive to timestamping
errors equal to an integer number of seconds, but it is very difficult
to imagine how such an error could arise in the ULTRASPEC data
acquisition system, and is sufficiently large that we would have
spotted it via our white-dwarf eclipse~monitoring,
e.g. \cite{marsh14}.

\subsection{Pipeline data reduction system}
\label{sec:pipeline}

ULTRASPEC can generate up to 0.6\,MB of data per second. In the course
of a typical night, therefore, it is possible to accumulate up to 25
GB of data, and up to 0.25 TB of data in the course of a typical
observing run. To handle these data rates, ULTRASPEC uses the same
dedicated pipeline reduction system used by
ULTRACAM\footnote{Available for download at
  http://deneb.astro.warwick.ac.uk/\newline
  phsaap/software/ultracam/html/index.html.}, written in C++, which
runs on a Linux PC or Mac located in the telescope control room and
connected to the instrument control PC via a dedicated Gbit local area
network (see Fig.~\ref{fig:das}).

As with ULTRACAM, ULTRASPEC data are stored in two files, one an XML
file containing a description of the data format, and the other a
single, large unformatted binary file containing all the raw data and
timestamps from a particular run on an object. This latter file may
contain thousands of individual data frames, each with its own
timestamp. The data reduction pipeline grabs these frames from the
hard disk by sending HTTP requests to a file server running on the
instrument control PC.

The data reduction pipeline has been designed to serve two apparently
conflicting purposes. Whilst observing, it acts as a quick-look data
reduction facility, with the ability to display images and generate
light curves in real time, even when running at the highest data rates
or frame rates. After observing, the pipeline acts as a fully-featured
photometry reduction package, including optimal extraction
\citep{naylor98}. To enable quick-look reduction whilst observing, the
pipeline keeps many of its parameters hidden to the user and allows
the few remaining parameters to be quickly skipped over to generate
images and light curves in as short a time as possible. Conversely,
when carefully reducing the data after a run, every single parameter
can be tweaked in order to maximise the SNR of the final
data.

\subsection{Readout modes}
\label{sec:modes}

\begin{figure*}
\centering
\includegraphics[width=14.2cm]{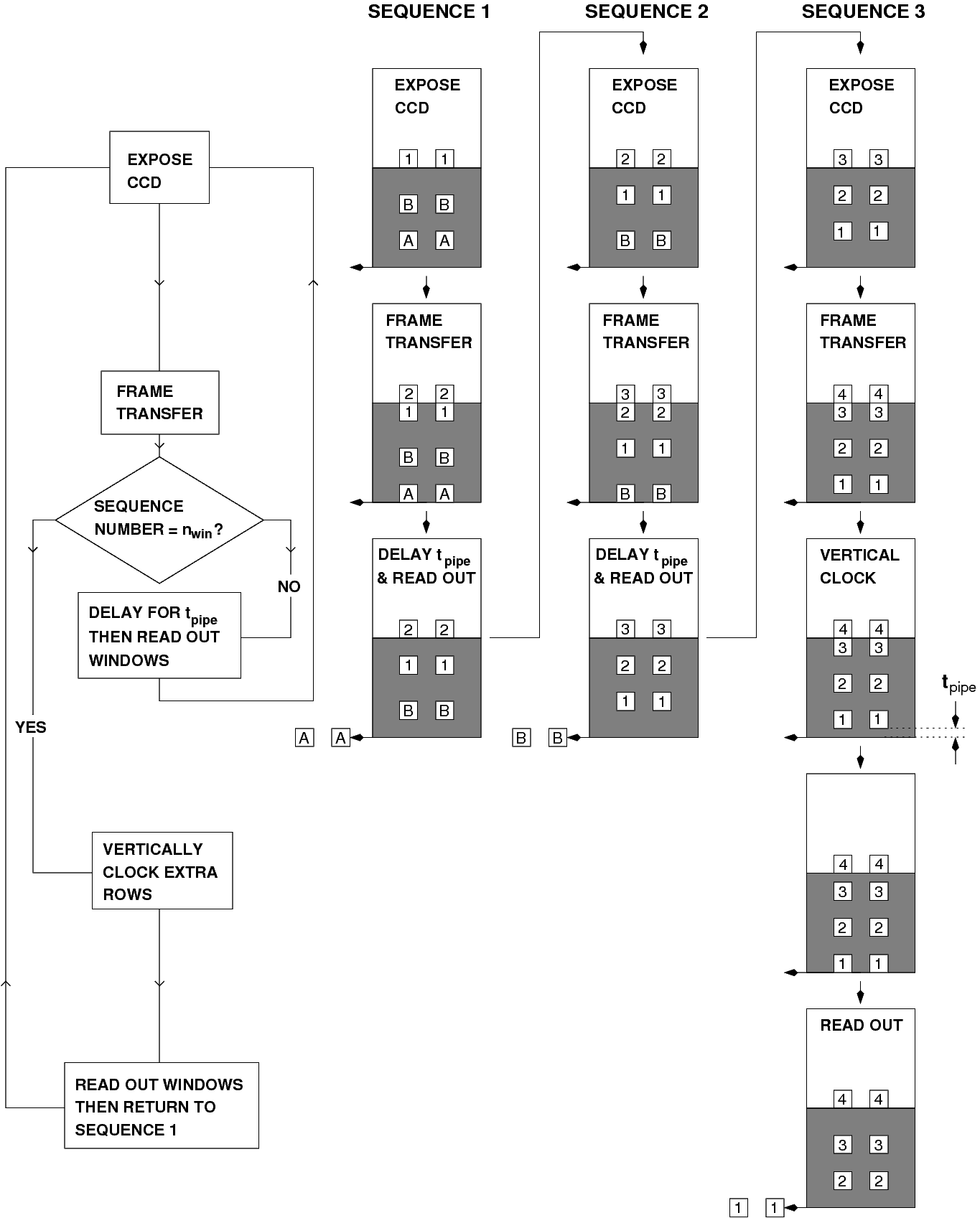}
\caption{Pictorial representation of the readout sequence in drift
  mode with three pairs of windows, i.e. $n_{win}=3$, as implemented
  in ULTRASPEC. In contrast to the standard windowed mode shown in
  \protect{Fig.~\ref{fig:frame}}, the two drift-mode windows must have
  the same vertical position and size. Exposed windows form a vertical
  stack in the storage area. The storage area has 1037 rows, implying
  that the vertical gaps between the windows can never all be the
  same. To maintain uniform exposure times and intervals between
  exposures, therefore, a pipeline delay, $t_{pipe}$, is added to
  sequences 1 and 2 (see Appendix~A of \protect\cite{dhillon07a} for
  further details). On completion of sequence 3, the cycle begins
  again by returning to sequence 1.}
\label{fig:drift_mode}
\end{figure*}

ULTRASPEC can be read out in a number of different modes, as
illustrated in Fig.~\ref{fig:frame}: full-frame mode, one-window,
two-windows, three-windows and four-windows, with each mode available
from both the normal and avalanche outputs.  By adjusting the position
of the telescope, the angle of the Nasmyth rotator, and the
positions/sizes of the CCD windows, it is possible to select multiple
comparison stars for differential photometry. The only restriction on
the window positions and sizes is that they must not overlap in the
vertical direction. On-chip binning factors ranging from 1 to 8 pixels
are also available.

Standard frame-transfer operation proceeds as follows. Once an
exposure is complete, the entire image area is shifted into the
storage area. This frame-transfer process is rapid, taking only
14.9\,ms. As soon as the image area has been shifted in this way, the
next exposure begins. Whilst exposing, the previous image in the
storage area is shifted row-by-row onto the serial register, dumping
any unwanted rows between windows, and then horizontally clocked to
one of the two outputs where it is digitised\footnote{For the purposes
  of this paper, the word {\em digitisation} shall refer to both the
  process of determining the charge content of a pixel via correlated
  double sampling and the subsequent digitisation of the charge using
  an analogue-to-digital converter.}, dumping any unwanted pixels
lying outside the defined windows. In other words, the previous frame
is being read out whilst the next frame is exposing, thereby reducing
the dead time between exposures to the time it takes to shift the
image into the storage area, i.e. 14.9\,ms.

It is important to note that ULTRASPEC has no shutter -- the fast
shifting of data from the image area to the storage area acts like an
electronic shutter, and is far faster than conventional mechanical
shutters. This does cause some problems, however, such as vertical
trails of star-light from bright stars, but these can be overcome in
some situations by the use of a focal-plane mask (see
Sect.~\ref{sec:mechanics}).

Setting an exposure time with ULTRASPEC is a more difficult procedure
than with a conventional non-frame-transfer CCD. This is because
ULTRASPEC attempts to frame as fast as it possibly can, i.e. it will
shift the image area into the storage area as soon as there is room in
the storage area to do so. Hence, the fastest exposure time is given
by the fastest time it takes to clear sufficient room in the storage
area, which in turn depends on the number, location, size and binning
factors of the windows in the image area, as well as the digitisation
speed (slow/medium/fast), all of which are variables in the ULTRASPEC
data acquisition system. To obtain an arbitrarily long exposure time
with ULTRASPEC, therefore, an {\em exposure delay} is added prior to
the frame transfer to allow photons to accumulate in the image area
for the required amount of time. Conversely, to obtain an arbitrarily
short exposure time with ULTRASPEC, it is necessary to set the
exposure delay to zero and adjust the window, binning and digitisation
parameters so that the system can frame at the required rate. As it
takes 14.9\,ms to vertically clock the entire image area into the
storage area, this provides a hard limit to the maximum frame rate of
68\,Hz, which results in essentially zero exposure time on source.  If
we instead set the maximum frame rate to the value which results in a
duty cycle (given by the exposure time divided by the sum of the
exposure and dead times) of 75\%, the maximum frame rate is~approximately~20~Hz.

If frame rates faster than about 20\,Hz are required, a completely
different readout strategy to that described above is required. For
this purpose, we developed {\em drift mode} for ULTRACAM, and we have
implemented it in ULTRASPEC. The readout sequence in drift mode is
shown pictorially in Fig.~\ref{fig:drift_mode} and described in detail
in Appendix~A of \cite{dhillon07a}. Briefly, in drift mode the windows
are positioned on the border between the image and storage areas and,
instead of vertically clocking the entire image area into the storage
area, only the window is clocked into the (top of) the storage area. A
number of such windows are hence present in the storage area at any
one time. This dramatically reduces the dead time, as now the dead
time between exposures is limited to the time it takes to clock only a
small window into the storage area, not the full frame. For example,
in the case of two windows of size 20$\times$20 pixels and binned
4$\times$4, it only takes 0.3\,ms to move into the storage area,
reducing the dead time by a factor of 50 and providing a frame rate of
200\,Hz with a duty cycle of 94\%. It is worth noting that at these
high frame rates it is the speed at which charge can be shifted along
the serial register (currently limited by the SDSU controller to
0.48\,$\mu$s/pixel for the normal output), rather than the
digitisation time, that limits the frame rate. With larger windows and
hence lower frame rates, the reverse is true. Drift mode has the
disadvantage that only two windows, instead of up to four, are
available. In addition, drift mode windows spend longer on the CCD,
thereby accumulating more dark current (albeit negligible in
ULTRASPEC) and, without the use of the focal plane mask, more sky
photons. Hence drift mode should only be used when the duty cycle in
non-drift mode becomes unacceptable, which typically occurs when frame
rates in excess of about 20 Hz are required. For example, we recently
used drift mode when performing lunar occultations with ULTRASPEC at
the TNT \citep{richichi14}.

Due to its complexity, drift mode only offers the possibility of two
windows, with no clearing between frames. In contrast to the standard
two-window mode shown in Fig.~\ref{fig:frame}, the two drift-mode
windows shown in Fig.~\ref{fig:drift_mode} must have the same vertical
position and size, although they can have different horizontal
sizes. The only difference between the drift mode implementation in
ULTRASPEC and ULTRACAM is that both windows must be read through the
same output (either normal or avalanche) in ULTRASPEC, whereas in
ULTRACAM there are two normal outputs which each read one drift-mode
window.

For some applications, e.g. when taking flat fields or observing
bright standard stars, it is desirable to use a full frame or one
large window and yet have short exposure times. To enable exposures
times of arbitrarily short length, therefore, ULTRASPEC also offers
users the option of turning CCD clearing on or off. When clearing is
turned on, the data in the image area is dumped prior to exposing for
the requested amount of time. This means that any photo-electrons
collected in the image area whilst the previous exposure is reading
out are discarded. The disadvantage of this mode is that the duty
cycle becomes very poor (1\% in the case of a 0.1\,s full-frame
exposure in slow readout mode).
 
\section{Performance on the TNT}
\label{sec:performance}

\begin{figure*}
\centering
\includegraphics[height=7.4cm,angle=90]{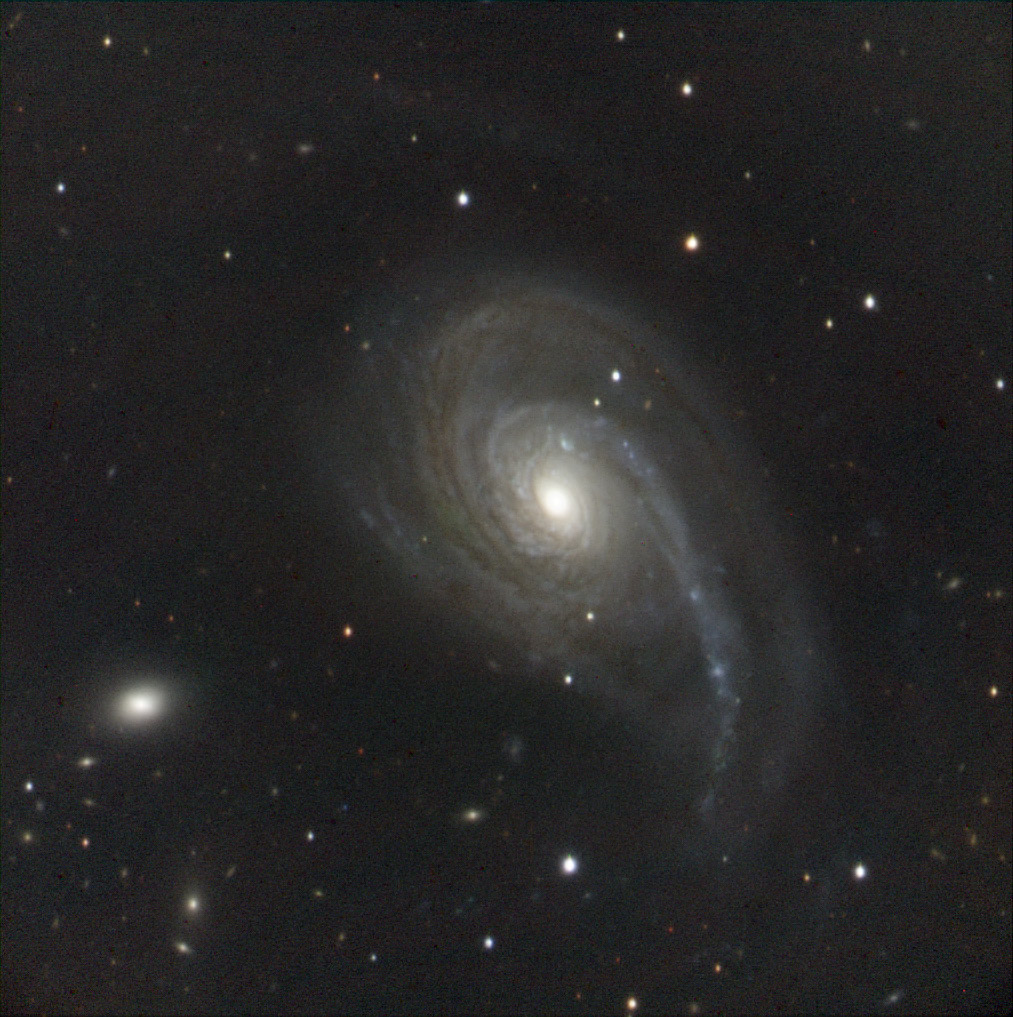}\hspace{0.5cm}\includegraphics[height=7.5cm,angle=0]{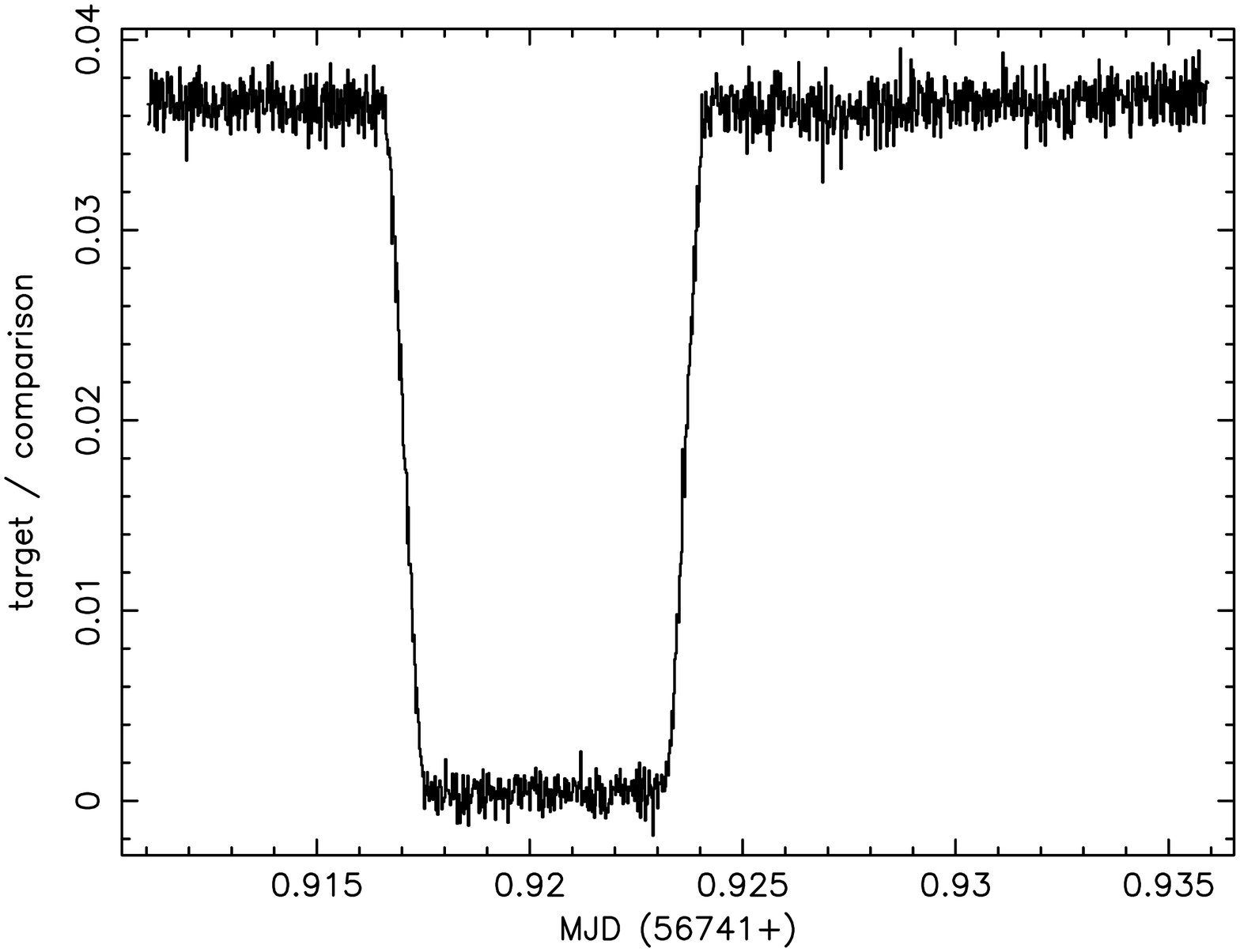}
\caption {Left: First light with ULTRASPEC on the TNT -- a full-frame
  image of the Nautilus Galaxy (NGC\,772). The image was taken on 2013
  November 5 and is composed of three separate exposures in $g'$, $r'$
  and $i'$, each of 180\,s duration. The field of view is 7.7', with
  North approximately to the top and East to the left. Right: An
  example light curve obtained with ULTRASPEC on the TNT, showing an
  eclipse of the white dwarf by the M-dwarf in the detached binary
  NN~Ser. The system has a magnitude of $r'=21.8/16.9$ in/out of
  eclipse \protect{\citep{parsons10}} and is believed to host two
  planets \protect{\citep{marsh14}}. The data were obtained in the KG5
  filter with an exposure time of 2\,s.}
\label{fig:first_light}
\end{figure*}

ULTRASPEC saw first light on the TNT on 2013 November 5 (see
Fig.~\ref{fig:first_light}) and the first scientific paper, reporting
on the drift-mode observation of lunar occultations, has recently been
accepted for publication \citep{richichi14}. In this section, we detail the
performance of ULTRASPEC on the TNT.

\subsection{Image quality}
\label{sec:image_quality}

In order to assess the image quality of ULTRASPEC on the TNT, we
observed the open cluster NGC\,6940. After carefully focussing the
telescope in each filter, we rapidly cycled through the $u'$, $g'$,
$r'$, $i'$, $z'$ filters, taking images of the cluster in each filter.

We astrometrically calibrated the images using {\em Astrometry.net}
\citep{lang10} and then determined the platescale in each filter,
finding the same value in all five filters: $0.452\pm0.001$'', as
designed. This meets one of the original requirements of the optical
design -- a plate scale independent of wavelength in the 330--1000\,nm
range. Moreover, we found no evidence for image shift as a function of
wavelength, with the stars occupying identical pixel positions in each
filter.

We then measured the FWHM of all of the stars in each image and
produced a map of image quality in each band. The seeing was
approximately 1.2'' during the tests, and we were able to confirm that
the FWHM of the stars in the central arcminute of the field of view
were $4.0\pm0.1$, $2.8\pm0.1$, $3.0\pm0.1$, $2.7\pm0.1$, $2.7\pm0.1$
pixels in $u'$, $g'$, $r'$, $i'$, $z'$, respectively, degrading by no
more than 10\% at the edge of the 7.7' field of view in each
filter. With the exception of the $u'$ band, therefore, the image
quality is a relatively insensitive function of wavelength and field
angle, as designed. Moreover, we regularly observed stellar FWHM below
2 pixels, i.e. seeing of below 0.9'', verifying that the ULTRASPEC
optics meet the requirement of providing seeing-limited stellar images
in median seeing conditions (0.9'') at the TNT.

Next, we investigated the level of vignetting by moving the cluster in
20 steps across the field of view, taking an image at each
position. We then measured the change in brightness of a number of
stars as a function of their position on the chip and found them to be
stable to within 2\% across the entire field of view, indicating that
there is no serious vignetting in our optics.

\subsection{Flexure}
\label{sec:flexure}

Whilst still observing the open cluster NGC\,6940, we rotated the
Nasmyth rotator through 360$^{\circ}$ and determined the rotator
centre, which we found lay ($-4$, $37$) pixels from the chip centre,
verifying the excellent mechanical alignment of ULTRASPEC.

We could not see any evidence for flexure of the ULTRASPEC mechanical
structure in the tracks of the stars whilst rotating, and a star
placed at the rotator centre moved by no more than 3 pixels,
indicating mechanical flexure of less than 39\,$\mu$m at the detector,
as designed.

\subsection{Sensitivity}
\label{sec:sensitivity}

\begin{figure}
\centering
\includegraphics[width=8.5cm,angle=0]{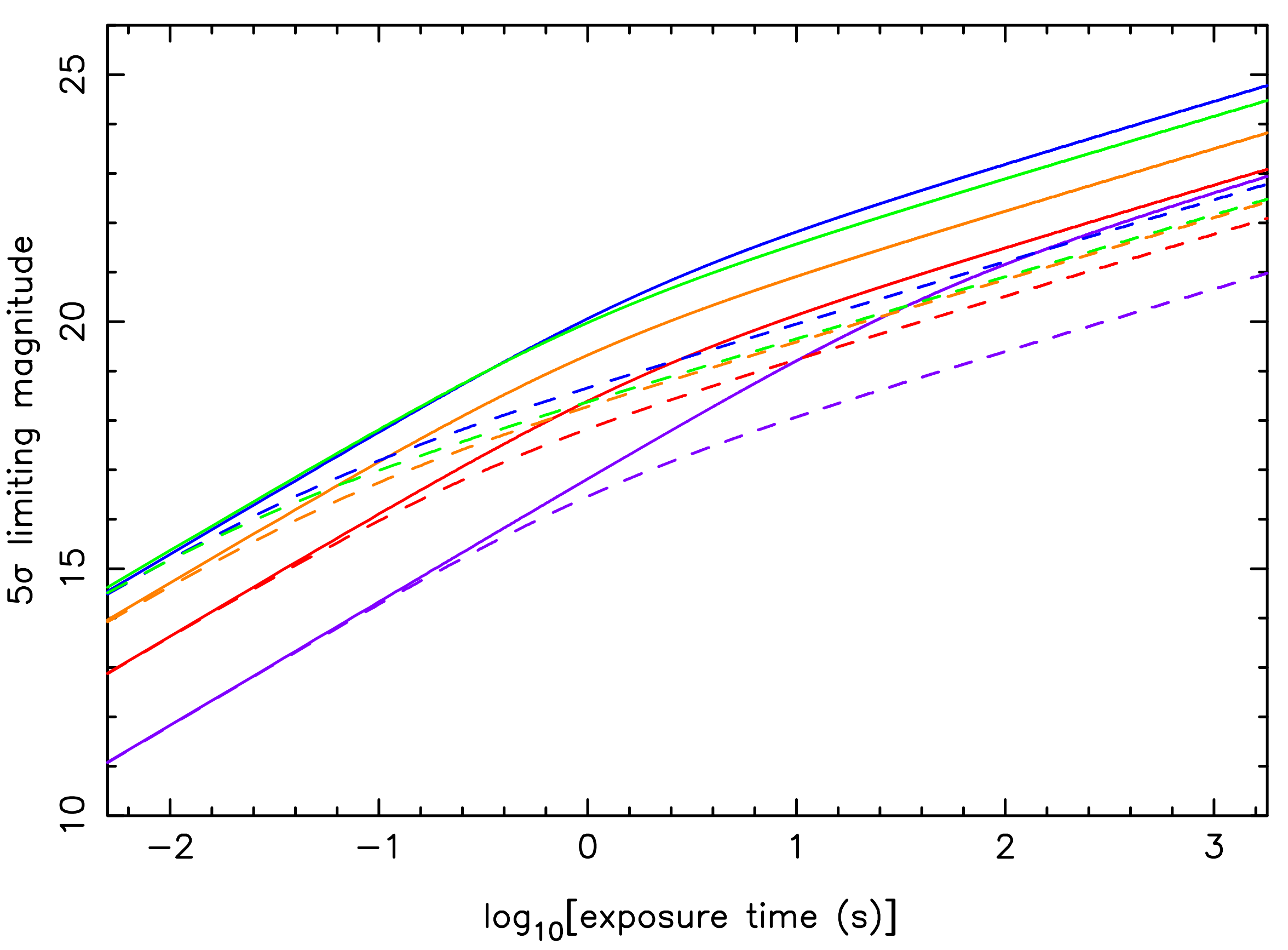}
\caption {Limiting magnitudes (5$\sigma$) of ULTRASPEC on the TNT as a
  function of exposure time. The purple, blue, green, orange and red
  curves show the results for the $u'$, $g'$, $r'$, $i'$ and $z'$
 filters, respectively. Solid lines show the results for dark time
  and dashed lines for bright time. The calculations assume that the
  normal output of the EMCCD is used, seeing of 1'', unity airmass, and
  no CCD binning.}
\label{fig:snr}
\end{figure}

By observing SDSS standard stars \citep{smith02} during commissioning
we were able to derive the zero points given in Tab.~\ref{tab:perf},
defined as the magnitude of a star above the atmosphere that gives 1
electron per second with ULTRASPEC on the TNT.  We also measured the
atmospheric extinction at Doi Inthanon using observations of
comparison stars obtained during multi-hour runs on variable stars. We
found values of $k_{g'}=0.20$ and $k_{r'}=0.10$, only $\sim10$\%\
worse than the extinction measured on the best nights at the
Observatorio del Roque de los Muchachos on La Palma
\citep{garcia-gil10}, for example. The above figures allow us to
estimate the throughput of the ULTRASPEC optics, i.e. just the lenses,
and not including the atmosphere, telescope, filter and CCD. The
throughputs are given in Tab.~\ref{tab:perf}, and to calculate them we
have assumed that the atmospheric extinction at Doi Inthanon is 10\%\
worse than on La Palma at all optical wavelengths.

\begin{table}
\centering
\caption{ULTRASPEC zero points and throughputs.}
\begin{tabular}{lccccc}
\hline\noalign{\smallskip}
 & $u'$ & $g'$ & $r'$ & $i'$ & $z'$ \\
\hline\noalign{\smallskip}
Zero point & 22.16 & 25.28 & 25.25 & 24.55 & 23.46 \\
Throughput (\%) & 12 & 51 & 61 & 45 & 37 \\
\hline\noalign{\smallskip}
\end{tabular}
\label{tab:perf}
\end{table}

\begin{figure*}
\centering
\includegraphics[width=9.9cm,angle=270]{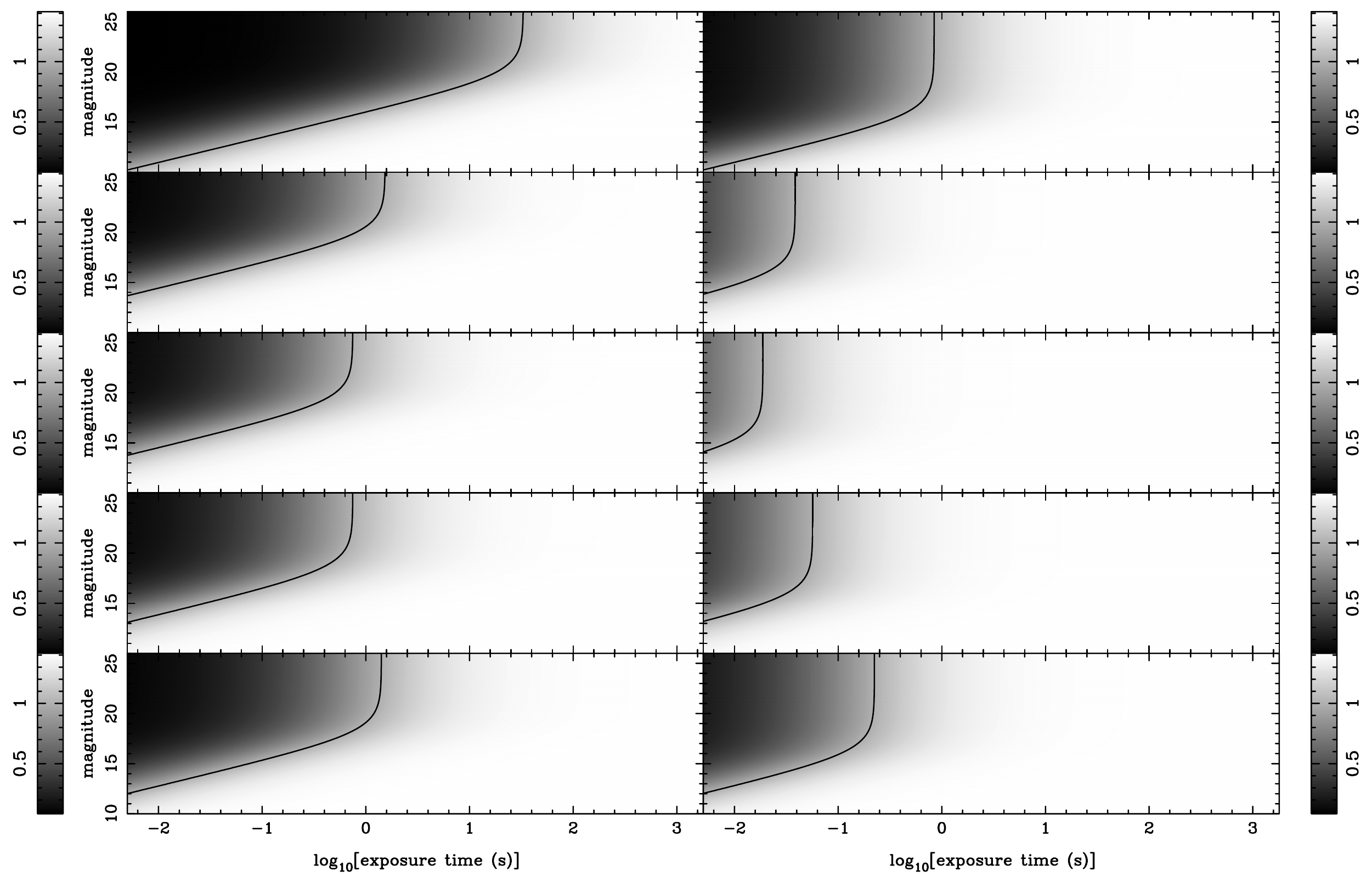}
\caption {Greyscale showing the ratio of the SNR obtained from the
  normal output of ULTRASPEC on the TNT to that obtained with the
  avalanche output, as a function of object magnitude and exposure
  time. The curved line in each panel indicates where both outputs
  would give identical SNR.  Hence the white regions to the right of
  this line indicate where the normal output would result in superior
  SNR and the black regions to the left of this line indicate where
  the avalanche output would result in superior SNR.  The left and
  right columns show the SNRs obtained in dark and bright time,
  respectively. The rows show, from top to bottom, the SNRs obtained
  with the $u'$, $g'$, $r'$, $i'$ and $z'$ filters. Black and white in
  each panel correspond to values of 0.085 and 1.414, respectively, as
  indicated in the greyscale wedges. The calculations assume seeing of
  1'', unity airmass and no CCD binning. For the normal output,
  readout noise of 2.3\,e$^-$ is assumed. For the avalanche output, we
  assume linear (or proportional) mode is used, where the readout
  noise is assumed to be zero and the QE of the EMCCD is effectively
  halved due to the presence of multiplication noise (see
  \protect{\citeauthor{tulloch11} (\citeyear{tulloch11}}) for
  details). As discussed by the latter authors, some of this effective
  QE loss can in principle be regained through photon counting, but it
  is difficult to avoid coincidence losses due to the high sky
  background when imaging at all but the highest frame rates in blue
  filters during dark time.}
\label{fig:norm_vs_aval}
\end{figure*}

Fig.~\ref{fig:snr} shows the 5$\sigma$ limiting magnitudes
achievable with ULTRASPEC as a function of exposure time and moon
brightness, calculated using the above zero points. The curves are
calculated assuming the normal output of the EMCCD is used, with a
readout noise 2.3\,e$^-$. As expected, $g'$ and $r'$ are by far the
most sensitive ULTRASPEC bands, able to achieve a limiting magnitude
of nearly 15 in 0.005\,s exposure times, and nearly magnitude 25 in
1800\,s exposure times.

Fig.~\ref{fig:norm_vs_aval} shows the ratio of the SNR obtained with
the normal output to that obtained with the avalanche output of
ULTRASPEC on the TNT, as a function of exposure time and object
magnitude. Values greater than unity lie to the right of the curved
line in each panel, and indicate where it is better to use the normal
output. Values less than unity lie to the left and indicate that the
avalanche output would be advantageous. It can be seen that in dark
time (left-hand panels), it is never worth using the avalanche output
for exposure times of longer than $\sim 1$\,s unless one is using the
$u'$ filter. This cut-off moves to $\sim 0.1$\,s in bright time. The
maximum benefit from using the avalanche output is obtained when
observing the faintest targets (magnitude $\gae 20$) at the highest frame
rates ($\gae 100$\,Hz) in dark time, when SNR gains of up to a
factor of 10 are available. Note that if the avalanche output is
erroneously used in the region lying to the right of the curved lines in
Fig.~\ref{fig:norm_vs_aval}, one loses a factor of $\sqrt{2}$ in SNR
due to the multiplication noise.

In practice, we almost never use the avalanche output of ULTRASPEC
when imaging on the TNT. There are a number of reasons for this.
First, the region of parameter space in which it is really
advantageous to use the avalanche output is quite small, as shown by
Fig.~\ref{fig:norm_vs_aval}, and it is extremely rare that we
require very fast observations of very faint targets (one exception
being observations of pulsars). Second, it is very easy to saturate
the multiplication register, which not only ruins the data but also
significantly reduces the life of the EMCCD (e.g. \citealt{evagora12}). Since we often observe
bright comparison stars simultaneously with faint target stars, great
caution is required. The CCD\,201-20 device used in ULTRASPEC has a
full-well capacity in the multiplication register of approximately
80\,000\,e$^-$. With an EM gain of $g_{A}\sim 1000$, this means that
it only requires 80\,e$^-$/pixel entering the multiplication register
to saturate it. Moreover, since $g_{A}$ is a mean value and some
pixels experience much higher amplification, it is safest to stay
below a much lower incident light level of, say, 20\,e$^-$/pixel. Even
using windows to mask out all but the faintest comparison stars does
not guarantee safety, as any bright stars outside these windows that
fall on the same CCD rows as the windows will pass through the
multiplication register, saturating it and reducing its lifetime.

\subsection{Reliability}
\label{sec:reliability}

ULTRASPEC has only two moving parts -- the focal-plane slide and the
filter wheel. This makes it an extremely reliable instrument. In the
observing season running from the start of November 2013 to the end of
April 2014 we estimate that we lost no more than 2\%\ of the time due
to technical problems with ULTRASPEC.

\subsection{Outstanding issues}
\label{sec:issues}

There are two main outstanding issues with ULTRASPEC on the TNT that
we hope to fix in time for the start of the 2014/2015 observing
season. The first is variable, high readout noise in the CCD,
typically up to 5\,e$^-$ instead of the 2.3\,e$^-$ we measured in the
lab. We believe this is due to the poor electrical earth of the
telescope, which is due to be improved during the summer of 2014 by
digging additional holes for copper earthing rods in the ground
surrounding the observatory.  The second is scattered light in the
central arcminute of ULTRASPEC, evident as a diffuse spot of emission
lying $\sim$5--10\%\ above the sky level in CCD images. We believe
this is due to poor telescope baffling around the M3 and M4 mirrors
and we hope to install better baffles to cure this problem during the
summer of 2014.

\section{Conclusions}
\label{sec:conclusions}

We have described the design and performance of ULTRASPEC, which has
just successfully completed its first observing season on the TNT. The
permanent presence of a high-speed imaging photometer on a telescope
of this size, and in this geographical location, provides us with a
powerful new tool to study compact objects of all classes, and to
perform rapid follow-up observations of transient astrophysical
events. It is our intention to continue operating ULTRASPEC on the TNT
for many years to come.

\section*{Acknowledgments}

We thank Edward Dunham for his referee's report, which improved the
clarity and usefulness of this paper. We would also like to thank the
staff of the Mechanical Workshop in the Department of Physics and
Astronomy at the University of Sheffield, the UK Astronomy Technology
Centre in Edinburgh, the National Astronomical Research Institute of
Thailand, and the European Southern Observatory at La Silla for their
valuable contributions to the ULTRASPEC project. We are grateful to
the European Commission (OPTICON), STFC, the Department of Physics of
the University of Warwick, the Faculty of Science at the University of
Sheffield, the Royal Society and the Leverhulme Trust for providing
the funds to build, operate and exploit ULTRASPEC.

\bibliographystyle{mn2e}
\bibliography{abbrev,refs}

\label{lastpage}

\end{document}